\documentclass[useAMS,usenatbib]{mn2e}
\usepackage{times}
\usepackage{graphicx, latexsym, amssymb, amscd, psfrag}
\usepackage{epsfig}

\title[Evidence for galaxies being pre-processed before accreted into clusters]{Evidence for galaxies
 being pre-processed before accreted into clusters}
\author[Mahajan]{Smriti Mahajan\thanks{E-mail:s.mahajan1@uq.edu.au}\\
 School of Mathematics and Physics, University of Queensland, Brisbane, QLD 4072, Australia}

\def\eg{{e.g. }}
\def\ka{{k$+$A }}

\begin{document}

\date{}

\pagerange{\pageref{firstpage}--\pageref{lastpage}} \pubyear{2012}
\maketitle

\label{firstpage}
%======================== ABSTRACT ==================================
\begin{abstract}
 
 I use the spectroscopic data for galaxies in and around 28 nearby ($0.02\leq z\leq0.06$) X-ray bright
 galaxy clusters, to show that the incidence of \ka (or post-starburst) galaxies (EW(H$\alpha$)$<2$\,\AA~in
 emission and EW(H$\delta$)$>3$\,\AA~in absorption) may be correlated with the accretion of small galaxy groups
 in clusters. 
 At $r<2r_{200}$, the \ka galaxies are found in regions of higher galaxy density relative
 to other cluster galaxies.  The \ka galaxies have a positively skewed distribution of absolute velocity,
 $|v_{los}|/\sigma_v$, where $v_{los}$ is the difference between the line-of-sight
 velocity of the galaxy and the cluster's mean, and $\sigma_v$ is the cluster's velocity dispersion.
 This distribution is statistically different from that of other cluster galaxies within $2r_{200}$,
 and in the same absolute velocity range. Moreover, 87\% of clusters in the sample studied here show statistically
 significant substructure in their velocity distribution, and 91.4\% of all \ka galaxies are found to be a part
 of one of these substructures with 4-10 members. These results suggest that star formation in
  these \ka galaxies is likely to have been quenched due to ``pre-processing"
 in a poor group-like environment before they are accreted into clusters. I also find a mild, but statistically
 significant trend in the fraction of \ka galaxies increasing with the temperature of the X-ray emitting
 gas in clusters.
 
\end{abstract}


%======================== Introduction===============================
\section{Introduction}

 Environment of galaxies play a key role in their evolution at all epochs. Recent work
 utilising data from the wide-field surveys such as the Sloan Digital Sky Survey (SDSS),
 have evidently shown that the galaxy number density and velocity fields play equally
 important roles in defining the immediate and the global environment of a galaxy
 \citep[\eg][]{balogh04,mahajan11,mmr}. Several authors
 have also speculated the role played by the large-scale supercluster environment in modulating
 galaxy properties, thereby extending the impact of environment to lower density regions
 such as galaxy groups \citep{rasmussen12}, outskirts of clusters \citep{haines11,mahajan12}, and
 even the inter-cluster filaments feeding
 them \citep[\eg][]{fadda08,porter08,pereira10,haines11}. Since galaxies span a continuous
 range in all observables such as broadband colours and star formation rate (SFR),
 most of these studies suggest that galaxies are likely being ``pre-processed" before they
 are accreted into clusters. 
  
 Galaxies presently transiting from an active, star-forming phase to passive
 evolution are called post-starburst\footnote{Although the terms
 \ka, E$+$A and post-starburst correspond to different spectral features depending upon the
 time since the last major starburst in a galaxy, they have been used interchangeably
 in the literature. In the context of this letter, a post-starburst galaxy as defined in \S\ref{data}
 is called a \ka galaxy unless stated otherwise. } (or k+A) galaxies \citep{dressler83}. 
 Spectra of such post-starburst galaxies are often characterised by negligible emission,
  but strong absorption in Balmer lines such as H$\alpha$
 and H$\delta$. While the former indicates absence of ongoing star formation,
 the latter feature is an indicator of the presence of late A and early B-type stars,
 implying that star formation in the galaxy was quenched 500--700\,Myr ago. Although
 post-starburst galaxies are found at all redshifts and in all environments, their frequency varies with
 environment, galaxy mass and redshift \citep{pogg09,peng10}.
 
 If star formation in a galaxy is quenched because of mechanisms related to the intra-cluster
 medium (ICM), the incidence of post-starburst galaxies should correlate with the properties of the
 ICM. The fraction of post-starburst galaxies has been reported to be as high as a quarter of the
 total population of cluster galaxies \citep{dressler83,dressler99,tran04} at intermediate
 redshifts ($0.3<z<0.6$). The trend continues to $z\sim0$, in the sense that in the nearby Universe
 most of the stars are forming in dwarf galaxies, and, star formation in a large fraction of such
 cluster dwarfs is found to be quenched in dense environments \citep{mahajan10}. 
 However, it is noteworthy that not only are different studies of post-starburst galaxies
 limited by different data attributes, but also employ different selection criteria for selecting
 post-starburst galaxies. Hence, a comparison of relative fraction of post-starburst galaxies
 amongst different environments in the same study is more insightful than a comparison
 between absolute fractions. 
 
 At $z\sim0$, post-starburst galaxies seem to prefer less-dense environments, similar to
 blue star-forming galaxies \citep{balogh05,hogg06,yan09}, suggesting that environmental
 mechanisms related to dense clusters are not the only phenomena responsible for quenching
 star formation in galaxies. Some post-starburst galaxies, however, may be a result of pre-processing 
 of galaxies before they are accreted into clusters \citep{hogg06,mahajan10}. As I will show below,
 while the immediate neighbourhood has an immediate impact on the properties of galaxies, the
  global environment can play an indirect role in the incidence of \ka galaxies.   
 
 In the next section I describe the datasets used in this work, followed by the analysis and results
 in \S\ref{s:frac} and \S\ref{s:env} respectively. The major findings are summarised in \S\ref{summary}.
 This analysis makes use of a $\Lambda$CDM concordance cosmological model
 with $H_0=70$\,km s$^{-1}$\,Mpc$^{-1}$, $\Omega_\Lambda=0.7$ and $\Omega_M=0.3$
 for calculating distances and magnitudes.

 \section{Data}
 \label{data}
 
 \subsection{Cluster parameters}
 
 I selected Abell clusters \citep{abell} at $0.02<z<0.06$ having X-ray
 temperature measurement from BAX (http://bax.ast.obs-mip.fr/). The
 spectroscopic data, complete to the absolute magnitude limit for each cluster
 corresponding to the SDSS spectroscopic catalogue's apparent magnitude limit of
 $r=17.77$ ($-19.26<M_r<-17.05$) was obtained from the SDSS (DR7).
 The velocity dispersion ($\sigma_v$) of these clusters was estimated using the $3\sigma$
 velocity dispersion clipping technique. The final sample comprises 28 clusters, each having
 at least $15$ galaxies in the cluster core ($r/r_{200}<0.5$) to evaluate $\sigma_v$.
 The  $\sigma_v$ was then used to estimate the radius
 of the cluster using the relation $r_{200}=\sqrt{3}\sigma_v/10H(z)$, where $H(z)$ is the Hubble
 constant at redshift $z$ \citep{carlberg97}.  
 
\begin{table}
\caption{Cluster sample}
\begin{tabular}{|c|c|c|c|c|c|}
\hline
Cluster &  RA & Dec   & $z$  & $\sigma_v$ & Tx (keV) \\ %\hline\hline
             &    (J2000)  &   (J2000)   &     & (km s$^{-1}$) & (1.5--2.4 keV) \\ \hline\hline
Abell   85 &  10.4613 &  -9.3067 & 0.056 &  710.40 &   6.45 \\
Abell  119 &  14.0500 &  -1.2575 & 0.044 &  744.19 &   5.62  \\
Abell  160 &  18.2117 &  15.5106 & 0.041 &  800.99 &   1.68  \\
Abell  168 &  18.7479 &   0.3664 & 0.045 &  501.82 &   2.56 \\
Abell  671 & 127.1292 &  30.4072 & 0.050 &  835.43 &   4.20   \\
Abell  779 & 139.9713 &  33.7639 & 0.024 &  559.22 &   2.97  \\
Abell  957 & 153.4913 &  -0.9175 & 0.046 &  760.08 &   2.75  \\
%Abell 1139 & 164.5204 &   1.5125 & 0.040 &  234.00 &   2.20  \\
Abell 1142 & 165.2333 &  10.5439 & 0.036 &  865.57 &   3.70  \\
%Abell 1177 & 167.3725 &  21.6914 & 0.032 &  303.21 &   1.48  \\
Abell 1185 & 167.7075 &  28.6753 & 0.031 &  542.75 &   3.90  \\
Abell 1291 & 173.0392 &  56.0386 & 0.053 & 1200.68 &   3.96  \\
Abell 1314 & 173.7183 &  49.0547 & 0.033 &  578.15 &   5.00  \\
Abell 1367 & 176.1283 &  19.8369 & 0.021 &  735.74 &   3.55  \\
Abell 1656 & 194.9583 &  27.9825 & 0.023 &  860.91 &   8.25  \\
Abell 1890 & 214.3925 &   8.1908 & 0.057 &  505.68 &   5.77  \\
Abell 1913 & 216.7167 &  16.6808 & 0.053 &  527.26 &   2.78  \\
Abell 1983 & 223.1842 &  16.7528 & 0.045 &  516.82 &   2.18  \\
Abell 1991 & 223.6267 &  18.6533 & 0.059 &  492.23 &   5.40  \\
Abell 2040 & 228.1879 &   7.4367 & 0.045 &  822.15 &   2.4  \\
Abell 2052 & 229.1892 &   7.0236 & 0.035 &  617.87 &   3.12  \\
Abell 2063 & 230.7571 &   8.6461 & 0.035 &  666.17 &   3.57  \\
Abell 2107 & 234.9504 &  21.7792 & 0.041 &  582.52 &   4.00  \\
Abell 2147 & 240.5708 &  15.9197 & 0.035 &  694.16 &   4.34  \\
Abell 2151 & 241.3133 &  17.7553 & 0.037 &  752.15 &   2.58  \\
Abell 2152 & 241.3429 &  16.4564 & 0.037 & 1986.56 &   2.41  \\
Abell 2197 & 247.0429 &  40.9131 & 0.030 &  677.81 &   2.21  \\
Abell 2199 & 247.1533 &  39.5300 & 0.030 &  726.92 &   3.99  \\
Abell 2399 & 329.3896 &  -7.7886 & 0.058 &  320.62 &   2.46  \\
Abell 2593 & 351.1288 &  14.6417 & 0.041 &  651.04 &   3.10  \\ \hline
\end{tabular}
\end{table}

  \subsection{Cluster galaxies}

 \begin{figure*}
\centering{
{\rotatebox{270}{\epsfig{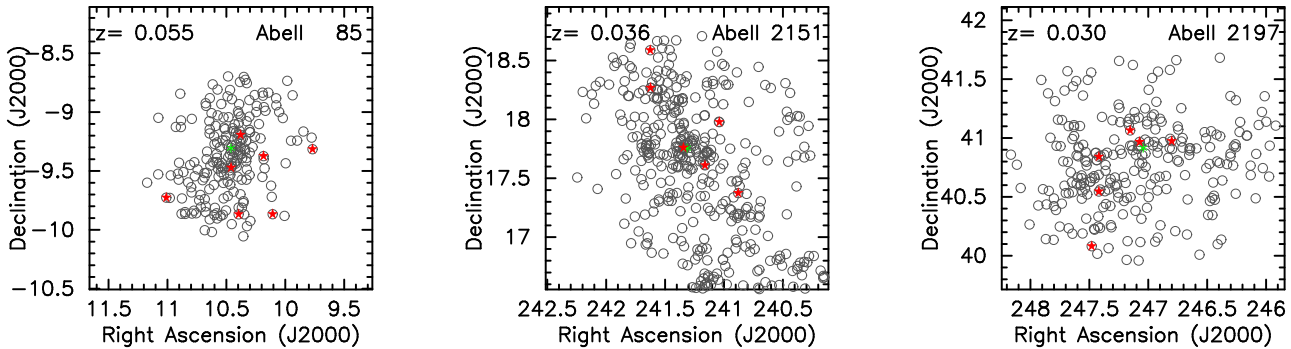}}}}
\caption{Some typical clusters in my sample showing the position of all
{\it{(grey)}} and \ka {\it{(red stars)}} galaxies within $2r_{200}$ of the cluster centre
 {\it{(green star)}}. }
\label{sky}
\end{figure*}
 
 In order to evaluate the cluster parameters mentioned above,
 I started with all galaxies within $\pm3,000$\,km s$^{-1}$ of the cluster's mean
 line-of-sight (l.o.s.) velocity, and within two degree radius. The final dataset comprises
 5,735 galaxies within $2r_{200}$ of the cluster centre, and having l.o.s. velocity
 such that the absolute velocity, $|v_{los}|\leq5\sigma_v$, where $v_{los}$ is the difference
 between the l.o.s. velocity of the galaxy, and the cluster's mean l.o.s. velocity. The factor of
 five is chosen to include all cluster galaxies, as well as those falling into clusters. 
 The position, $\sigma_v$ and Tx for all clusters are listed in Table~1. 
      
 In this letter a galaxy is defined as post-starburst (or k$+$A), if 
 its equivalent width (EW) in H$\alpha<2$\AA~in emission,
 and EW(H$\delta$)$>3$\AA~in absorption \citep[\eg][]{pogg09}. This gives a total of $165$
 \ka galaxies within $2r_{200}$ and $|v_{los}|/\sigma_v\leq5$ in $28$ X-ray bright clusters
 ($0.02<z<0.06$). Distribution of these galaxies in some typical clusters from the sample
 studied here are shown in Figure~\ref{sky}. 
 
 \section{\ka galaxy fraction and ICM temperature}
 \label{s:frac}

\begin{figure}
\centering{
{\rotatebox{270}{\epsfig{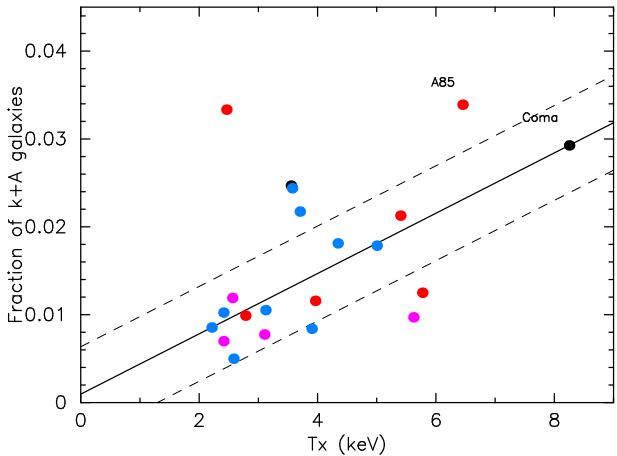}}}}
\caption{The fraction of \ka galaxies as a function of the temperature of the X-ray emitting
gas in cluster. Clusters are colour-coded by redshift, in increasing order {\it{black, blue, pink
 and red}}, for $0.02\leq z \leq0.6$ at an interval of $0.01$ respectively. The {\it{solid black}}
  line shows the mean absolute deviation (MAD) relation for all
 clusters, with $\pm1\sigma$ deviation in the fit shown by the {\it{black dashed}} line. The fraction is
 calculated for for galaxies brighter than $M_r<-19.35$, which is the magnitude limit
 for the SDSS spectroscopic catalogue at $z=0.06$. }
\label{frac}
\end{figure}
 
 Nearby clusters offer a unique opportunity to study galaxies in a `continuous' range
 of environments. Hence, whether galaxies are being transformed due to the environmental effects of the
 ICM, or prior to encountering it, they can be caught in the act. In order to test the hypothesis
 that the incidence of \ka galaxies is related to the cluster environment, I chose the most extreme
 environments in the nearby Universe by using the temperature of the X-ray emitting gas in the cluster
 as a proxy. This allowed me to sample cluster galaxies down to at least $M^*+1.48$ \citep{blanton01}
 using the SDSS (DR7) spectroscopic galaxy catalogue. These data were used to
 evaluate the fraction of \ka galaxies using galaxies ($M_r<-19.35$) within $2r_{200}$ and $|v_{los}|/\sigma_v\leq5$.
 The magnitude limit is chosen so as to evaluate the \ka fraction with the same lower luminosity bound for
  clusters at all redshifts. Figure~\ref{frac} shows the fraction of \ka galaxies as a function of the X-ray
  temperature of the ICM. 
  
 Figure~\ref{frac} shows a mild trend for an increase in the \ka fraction with Tx, with product moment  
 correlation coefficient, $r=0.481$ ($P=0.023$).
 The two hottest systems in the sample (Abell\,85 and Coma), are also known to have substantial
 substructure in galaxy distribution \citep{porter05,mahajan10,mahajan11},
 and, X-ray data \citep{neumann03,bravo09}.  Simulations by \citet{bekki10} showed
 that the post-starburst galaxies may be related to the presence of substructure in galaxy clusters.
 But with only a couple of data points at Tx$>6$\,keV, no substantial conclusions can be drawn to
 defend (or reject) any correlation between the fraction of \ka galaxies and Tx.

 \section{Local environment of \ka galaxies}
\label{s:env}
 
\begin{figure}
\centering{
{\rotatebox{270}{\epsfig{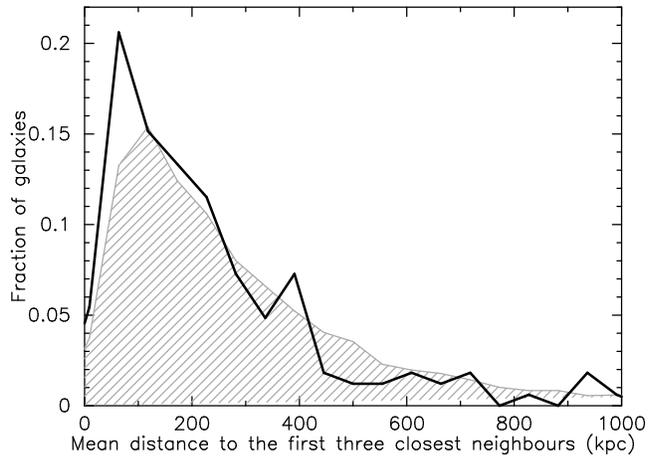}}}}
\caption{The distribution of the mean distance to the three closest neighbours
 (within $\pm2,000$\,km s$^{-1}$) for all {\it{(grey hatched)}} and \ka {\it{(solid black)}}
 cluster galaxies are statistically different. Together with
 the results from Figure~\ref{vel}, this show that most of the \ka galaxies near rich clusters
 are being pre-processed in poor group environments before they mingle with the cluster
 population. }
\label{d3}
\end{figure}

\begin{figure}
\centering{
{\rotatebox{270}{\epsfig{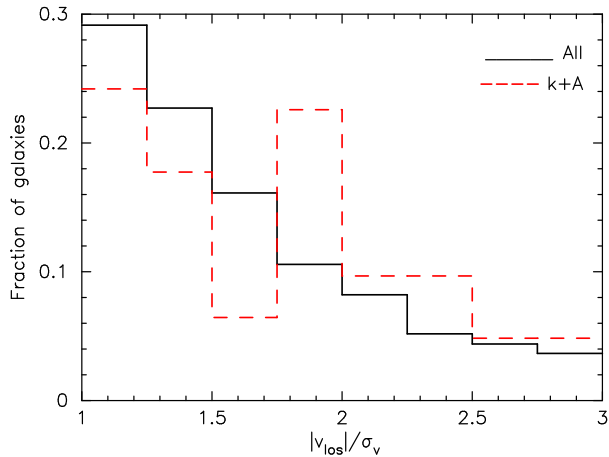}}}}
\caption{The distribution of absolute l.o.s. velocity for all
 {\it{(solid black)}} and \ka {\it{(red dashed)}} galaxies within $2r_{200}$ of the cluster centre 
 and $1\leq|v_{los}|/\sigma_v\leq3$. These distributions are statistically
 different (see text), confirming that most \ka galaxies are likely to be falling into
 clusters. }
\label{vel}
\end{figure}

 Figure~\ref{d3} shows the distribution of the mean distance to the three closest neighbours
 of a galaxy ($D_3$; chosen to be within a redshift slice, $\Delta z=\pm2,000$\,km s$^{-1}$
 around the galaxy), for the samples mentioned above. The Kolmogorov-Smirnov (KS) test in favour of
 the hypothesis that both samples are drawn from the same parent population
 yields a probability of $0.0190$ with median $D_3=248.9$ and $199.5$\,kpc, and maxima for the distributions
 being $118.1$ and $63.6$\,kpc, for all and \ka
 cluster galaxies, respectively. In order to test if the discrepancy in the relative local densities
 of the two galaxy samples is dependent on $\Delta z$ chosen for evaluating $D_3$, I re-calculated
 median $D_3$ including neighbours only within $\Delta z=\pm1,000$\,km s$^{-1}$ to
 be $294.4$ and $257.3$\,kpc for
 all and \ka galaxies respectively. This suggests that the \ka galaxies are found in locally over-dense
 regions within clusters.  Ideally, such an analysis should be performed in small radial bins 
 in order to disentangle the impact of local over-density from that of the large-scale environment
 on \ka galaxies, but the rarity of \ka galaxies in low-redshift clusters only permits a stacked analysis. 
 
  Figure~\ref{vel} shows the distribution of absolute velocity of all cluster galaxies within $2r_{200}$
 and $1\leq|v_{los}|/\sigma_v\leq3$, along with the \ka galaxies in the same radial and absolute velocity
 range. The exclusion of galaxies with $|v_{los}|/\sigma_v<1$ allows me to focus on the population
 of mostly non-virialized galaxies. 
 The KS statistic probability in favour of the hypothesis that
 these two distributions are drawn from the same parent population is $0.0098$, suggesting
 that a non-negligible fraction of \ka galaxies may have accreted into the clusters recently. 
 
  \citet{pogg09} found \ka galaxies to be more frequent in low velocity dispersion galaxy groups
 (200--400\,km s$^{-1}$) and massive clusters \citep[$0.4<z<0.8$; also see][]{vergani10}, relative
 to the field and rich groups. In the Coma supercluster ($z=0.03$), \citet{mahajan11} found that
  \ka galaxies mainly occurred on cluster outskirts and galaxy groups. 
  I applied the ``$\kappa$-test" devised by \citet{colless96} to test for localized variations in the velocity
 distribution of galaxies in each cluster. This test compares the velocity distribution of a group of $N$
 nearest neighbours of each galaxy to the cluster's mean. The significance of $\kappa_N$ is
 estimated by Monte Carlo simulations in which velocities of cluster galaxies are shuffled randomly.
 The test shows that 24/28 clusters in this sample are at least 95\% likely to have substructures
 with 4-10 member galaxies. Furthermore, 91.4\% of all \ka galaxies are found to be a part of one of
 these substructures (P($\kappa_N>\kappa_{obs}$)=0.03\%).   
 Together, these results suggest that the \ka galaxies {\it{locally}}
 reside in marginally denser environment such as poor galaxy groups, and their presence
 may be correlated with accretion of such groups into clusters. 
  
 Quenching of star formation in \ka galaxies may be a result of the ``pre-processing"
 happening in these small groups assembling into larger clusters. \citet{struck06} suggested that for a
 poor group falling into a rich cluster with comparable halo core density, the galaxy density for
 group galaxies can increase by an order of magnitude and their collision rate by a factor of about 100
 (density squared), thereby increasing the probability of a burst of star formation in the process. 
 Such bursting galaxies will eventually be observed as the post-starburst population such as that
 studied here. 

 Simulation of a galaxy group merging with a cluster four times more massive than itself, shows
 that starburst can be triggered in gas-rich galaxies due to the external pressure of the
 ICM \citep{bekki10}. These starburst galaxies will eventually be transformed into
 a population of post-starburst galaxies. Bekki et al. argued that after the structure has finally
 relaxed ($\sim 5$\,Gyr), such galaxies will not show any preferred spatial distribution.
 For the sample studied here, I find that although the absolute velocity distribution of the \ka galaxies is
 statistically different from other cluster galaxies in the same absolute velocity regime, they do not
 show any preferred spatial distribution relative to the cluster centre (the KS test probability in favour of the
 hypothesis is $0.0927$ for galaxies within $2r_{200}$).
 If we assume several small groups merging into present day galaxy clusters, such a random
 distribution of group galaxies being pre-processed is expected.  
 
 However, several authors \citep{hogg06, delucia09, yan09} have argued against any correlation between the
 incidence of post-starburst galaxies and their global environment, suggesting that the post-starburst
 phase is not a dominant channel for converting the blue, star-forming galaxies
 to passively evolving red cluster galaxies. This letter presents a new angle to this dichotomy,
 by evidently showing that the incidence of \ka galaxies in nearby rich clusters is correlated
 with {\it{locally}} over-dense regions within clusters. Since most of the \ka galaxies in these clusters
 belong to group scale substructures, it is likely that the \ka galaxies are being pre-processed in such
 galaxy groups during their accretion into the cluster. Evidence for such pre-processing has also been
 reported by \citet{jaffe12} in galaxy groups falling into a cluster at $z\sim0.2$. Using optical and
 ultraviolet data for galaxies in optically-selected groups ($z\sim0.06$),
 \citet{rasmussen12} showed that not only SFR, but also SFR/$M^*$ is quenched in group galaxies.
 Moreover, these authors found the effect to be strongest for low-mass ($10^7-10^9 M_\odot$)
 galaxies.

 \begin{figure}
\centering{
{\rotatebox{270}{\epsfig{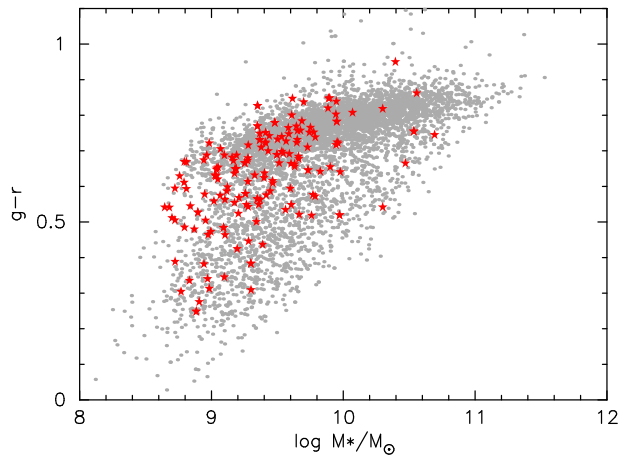}}}}
\caption{Distribution of all {\it{(grey points)}} and \ka cluster galaxies {\it{(red stars)}} in
 the $(g-r)-M^*$ space shows that most \ka galaxies in nearby rich clusters are spread over
 a large colour range. However, most of the \ka galaxies are less than $10^{10}\!M_\odot$
 in stellar mass.   }
\label{mass}
\end{figure}

 Figure~\ref{mass} shows the distribution of all cluster galaxies, and, \ka
 galaxies in the $(g-r)-M^*$ space. The stellar masses are estimated using the relation
 given by \citet{bell03}. This figure shows that 
 \ka galaxies around clusters are mostly less than $10^{10}\!M_\odot$ in stellar mass,
 but span almost the entire colour range populated by all cluster galaxies taken together. This result
 is in agreement with previous studies which showed that at $z\sim0$ post-starburst\footnote{Different authors
 have used different combinations of spectroscopic indices such as the EW of balmer lines and
 the 4000\AA~break, to classify post-starburst galaxies.}
 galaxies are more frequent among low-mass galaxies \citep[\eg][also see Rasmussen et al.
 2012]{mahajan10,mahajan11,geha12}. This may happen because (i) most of the star formation 
 at the present epoch is happening in low-mass galaxies, which are quenched during accretion
 into clusters, or, (ii) only low-mass galaxies go through a post-starburst phase when quenched. 
 Advanced simulations including better semi-analytic models of galaxy formation, environmental
 processes and feedback recipes are required to throw more light on this subject. Incidentally,
 some work has already begun in this direction \citep[\eg][]{delucia12}.
   
\section{Summary}
\label{summary}

 I have used the X-ray and optical spectroscopic data for nearby ($0.02\leq z \leq0.06$)
 galaxy clusters to show a mild trend of an increase in the incidence of \ka galaxies
 with Tx. However, a cluster sample uniformly spanning the whole range in Tx is required
 to confirm this. Such a correlation, if found, will confirm that the ICM plays a
 critical role in quenching star formation in infalling galaxies.

 For the X-ray bright clusters studied here, although \ka galaxies show only
 a marginal (statistically insignificant) preference for higher clustercentric distances, their absolute velocity
 distribution is different from other cluster galaxies ($<2r_{200}$, $1\leq|v_{los}|/\sigma_v\leq3$).
 These \ka galaxies are also found in locally over-dense regions relative to other cluster galaxies.
  86\% of clusters in this sample show statistically significant substructure on group scales
 in velocity distribution, with 91.4\% of all \ka galaxies belonging to one of these groups with 4-10 members.
 Together, these results
 suggest that \ka galaxies are likely to be members of poor galaxy groups, whose star formation may have been 
 quenched as a result of pre-processing during accretion into clusters.
  
 These results favour the argument that the SFR in galaxies falling
 into clusters may be enhanced on the cluster outskirts by a burst of star formation
 triggered due to galaxy-galaxy interactions \citep[\eg][and references therein]{fadda08,porter08,mahajan12}.
 However, it remains debatable whether the global environment plays any significant role in accelerating these
 interactions when the group encounters the cluster potential \citep{struck06,bekki10,jaffe12}, or, the group
 environment self-sufficiently aids the evolution of galaxies \citep{pereira10,rasmussen12},
 irrespective of the large-scale environment.

\section{Acknowledgements}

Funding for the SDSS and SDSS-II has been provided by the Alfred P. Sloan Foundation,
 the Participating Institutions, the National Science Foundation, the U.S. Department of Energy,
 the National Aeronautics and Space Administration, the Japanese Monbukagakusho, the 
 Max Planck Society, and the Higher Education Funding Council for England. The SDSS Web
 Site is http://www.sdss.org/. The SDSS is managed by the Astrophysical Research Consortium
 for the Participating Institutions. 
 
 This research has made use of the X-Rays Clusters Database (BAX)
 which is operated by the Laboratoire d'Astrophysique de Tarbes-Toulouse (LATT),
 under contract with the Centre National d'Etudes Spatiales (CNES).
 
 A copy of the fortran code used for evaluating the ``Kappa"-statistic was provided by Prof. Michael
 J. Drinkwater, University of Queensland. I convey sincere gratitude to the anonymous referee whose
 feedback greatly helped in improving the content of this paper. 
 
 \label{lastpage}

%%%%%%%%%%%%%% Bibiliography %%%%%%%%%%%%%%%%%%%%%%%%%%%%%%%

\end{document}